\documentclass[10pt]{article}  
\usepackage{graphicx}
\usepackage{amssymb}
\usepackage{color}
\input epsf
\parskip=\medskipamount
   \def\CaH{{\cal H}}

   \def\Ga{\Gamma}

\def\te{\theta}   
\def\om{\omega}   
      
\def\IB{\relax{\rm l\kern-.18 em B}}
\def\IC{\relax{\rm l\kern-.50 em C}}
\def\IE{\relax{\rm l\kern-.12 em E}}
\def\IK{\relax{\rm l\kern-.18 em K}}
\def\IL{\relax{\rm I\kern-.18 em L}}
\def\IN{\relax{\rm I\kern-.18 em N}}
\def\IR{\relax{\rm I\kern-.18 em R}}

\def\smallonehalf{\frac{{}_1}{{}^2}}
 
\def\\{\hfill\break}
\def\smallonehalf{\frac{{}_1}{{}^2}}

\def\Re{\mathop{\rm Re}\nolimits}
\def\Im{\mathop{\rm Im}\nolimits}

\font\tenfrak=eufm10  \font\sevenfrak=eufm7  \font\fivefrak=eufm5
\newfam\frakfam
\textfont\frakfam=\tenfrak
\scriptfont\frakfam=\sevenfrak
\scriptscriptfont\frakfam=\fivefrak

\newtheorem{proposition}{Proposition}

\def\frac#1#2{{#1\over #2}}
\def\fracpd#1#2{\frac{\partial #1}{\partial #2}}
\def\ptos{\leaders\hbox to 2mm{\hfil{.}\hfil}\hfill}

\begin{document}

\title{Superintegrable systems with a position dependent mass : 
Kepler-related  and Oscillator-related  systems }

\author{ Manuel F. Ra\~nada  \\ [3pt]
{\sl Dep. de F\'{\i}sica Te\'orica and IUMA } \\
  {\sl Universidad de Zaragoza, 50009 Zaragoza, Spain}     }
\date{May, 4  2016}
\maketitle 
 

\begin{abstract}
The superintegrability of two-dimensional Hamiltonians with a position dependent mass (pdm)  is studied (the kinetic term contains a factor $m$ that depends of the radial coordinate). 
First, the properties of Killing vectors are studied and the associated Noether momenta are obtained.
Then the existence of several families of superintegrable Hamiltonians is proved and the quadratic integrals of motion are explicitly obtained. 
These families include, as particular cases, some  systems previously obtained making use of different approaches. 
We also relate the superintegrability of some of these pdm systems with the existence of complex functions 
endowed with  interesting Poisson bracket properties. 
Finally the relation of these pdm Hamiltonians with the Euclidean Kepler problem and with the Euclidean harmonic oscillator is analyzed. 

\end{abstract}

\begin{quote}

{\smallskip}
{\sl Keywords:}{\enskip} Integrability ; Superintegrability ;    Killing vectors ; 
Position dependent mass ; Quadratic  constants of motion ;    Complex functions ; 

{\smallskip}
{\sl Running title:}{\enskip}
Superintegrable systems with a position dependent mass

{\smallskip}
AMS classification:  37J35 ; 70H06

{\smallskip}
PACS numbers:  02.30.Ik ; 05.45.-a ; 45.20.Jj
\end{quote}

\vfill
\footnoterule{\small
\begin{quote}
  {\tt E-mail: {mfran@unizar.es}  }
\end{quote}
}

\newpage


\section{Introduction}

It is known that some Liouville integrable systems, as the harmonic oscillator or the Kepler problem, 
admit more constants of motion than degrees of freedom; they are called superintegrable.
Therefore, a Hamiltonian $H$ with two degrees of freedom is said to be integrable 
if it admits an integral of motion $J_2$ in addition to the Hamiltonian, and superintegrable if 
it admits two  integrals of motion, $J_1$ and $J_2$,  that Poisson commute 
and a third independent integral $J_3$. 
The  integral $J_3$ has vanishing Poisson bracket with $H$ but not necessarily with $J_1$ and $J_2$.

The mass $m$ has been traditionally considered as a constant in the theory of physical systems  admitting a Hamiltonian description. 
A consequence of this is that the study of superintegrable systems has been mainly focused on two and three degrees of freedom natural Hamiltonian systems  (that is, kinetic term plus a potential) with a constant mass; in geometric terms this means that the configuration space $Q$ is an Euclidean space  or a constant curved space  (spherical or hyperbolic). 
Nevertheless, in these last years the interest for the study of systems with a position dependent mass has become a matter of great interest and has attracted a lot of attention to many authors. 
It seems therefore natural to enlarge the study of superintegrability to include systems with a position dependent mass.

It is known that the Liouville formalism characterize the Hamiltonians that are integrable but it does not provide a method  for obtaining the constants of motion; therefore it has been necessary to carry out several different methods for searching integrals of motion (Noether symmetries, Hidden symmetries, Lax pairs formalism, bi-Hamiltonian structures, etc). 
In a recent paper Szuminski et al studied \cite{SzumiMacPrz15} families of Hamiltonians of the form 
$$
  H_{nk}  =   {\smallonehalf}\,r^{n-k}\,\Bigl( p_r^2 +  \frac{p_{\phi}^2}{r^2} \Bigr) +  r^n U(\phi) \,,
$$
($n$ y $k$ are integers) and then, making use of some previous results of  Morales-Ruiz and Ramis  related with the differential Galois group of variational equations \cite{MoralesRuiz99, MoralesRR01, Nakagawa01}, they derive necessary conditions for the integrability of such systems. 
Then using some rather involved mathematics (related with the hypergeometric differential equation) they arrive to a certain number of Hamiltonians and prove that four of them, given by 
$$
 \CaH_1 =  {\smallonehalf}\,r^6\,\Bigl( p_r^2 +  \frac{p_{\phi}^2}{r^2} \Bigr) -  r \cos\phi \,,{\quad} (n=1, k=-5)
$$
$$
  \CaH_2  = {\smallonehalf}\,\frac{1}{r^2}\,\Bigl( p_r^2 +  \frac{p_{\phi}^2}{r^2} \Bigr) -  \frac{1}{r}\, \cos\phi \,,{\quad} (n=-1, k=1)$$
$$
  \CaH_3  =   {\smallonehalf}\,r^4\,\Bigl( p_r^2 +  \frac{p_{\phi}^2}{r^2} \Bigr)  -  \frac{1}{r}\, \cos\phi \,,{\quad} (n=-1, k=-5)
$$
$$
  \CaH_4  =   {\smallonehalf}\Bigl(p_r^2 +  \frac{p_{\phi}^2}{r^2} \Bigr)  -  r\, \cos\phi ,,{\quad} (n=1, k=1)
$$
are superintegrable (two independent constants in addition to the Hamiltonian). 
The fourth Hamiltonian is in fact a rather simple Euclidean system but the other three are really interesting and deserve be studied with detail.

In this paper we will study the existence of superintegrability and we will construct the constants of motion using as starting point the properties of the Killing vectors.

We recall that a Killing vector field $X$ in a Riemannian manifold $(M,g)$,  is the (infinitesimal) generator of a symmetry of the metric $g$ (that is, $X$ is a generator of isometries); 
in geometric terms $X$ must be solution of  the equation
$
  {\cal L}_Xg = 0
$ 
where ${\cal L}_X$ denotes the Lie derivative. 
If $M$ is of dimension $n$ then the metric admits at most $d = {\smallonehalf} n (n+1)$ linearly independent Killing vector fields (constant curvature spaces admit the maximum number; for example if $M$ is the Euclidean plane $M=\IE^2$ then $d=3$). 

 If the configuration space of a system is a  Riemannian manifold $(Q,g)$ then $g$ determines a kinetic Lagrangian $L_g=T_g = {\smallonehalf}g_{ij}\dot{x}^i\dot{x}^j$ such that  the associated motion is just the geodesic motion, and the Killing vectors of $(Q,g)$ determine the constants of motion for the geodesic trajectories (the so-called Noether momenta). 
In most of cases the addition of a potential $V(x)$ to the kinetic Lagrangian $L_g$ destroys these first integrals but, in some cases, the new system admits first integrals of second order in the momenta whose quadratic terms are  determined by  Killing tensors of valence $p=2$ built from Killing vectors. 
We recall that Killing tensor ${\bf K}$ of valence   $p$ defined in a Riemannian manifold $(M, g)$  is a symmetric $(p, 0)$ tensor satisfying the Killing tensor equation 
$$
  [{\bf K},g]_S = 0
$$
where $[\cdot,\cdot]_S$ denotes the Schouten bracket (bilinear operator representing the natural generalization of the Lie bracket of vector fields)   \cite{ChadGMc06, HorMcLSm09}.  
In the case $p=2$ the Killing tensor ${\bf K}$ determines a homogeneous quadratic function $F_K = K^{ij}p_i p_j$ and then the Killing equation can be rewritten as the vanishing of the Poisson bracket of two functions 
$$
   \{ K^{ij}p_i p_j  \,,\,  g^{ij}p_i p_j\} = 0 \,.  
$$
This means that the function $F_K$  is a first integral of the geodesic flow determined by the Hamiltonian ${\CaH}= (1/2)g^{ij}p_i p_j$. 
From a practical viewpoint this means that quadratic term of the integrals of the Hamiltonian $H=T_g+V$ can be expressed as a sum of products of the Noether momenta.

The three Hamiltonians $\CaH_j$, $j=1,2,3$, studied in \cite{SzumiMacPrz15} can be considered as Hamiltonians with position dependent masses (pdm) $m=1/r^6$, $m= r^2$, and $m=1/r^4$, respectively.
In geometric terms this means that they are defined in non-Euclidean spaces. 

The following three points summarize the contents of this paper.

\begin{itemize}
\item  We will study the existence of superintegrable systems with a position dependent mass (pdm) of the form $m_n=r^{2n}$  using the geometric formalism as an approach. We first  obtain the Killing vectors for the corresponding metrics (that are conformal metrics) and then we obtain the expressions of the Noether momenta. The following step is the  obtainment of the quadratic integrals. 

\item  In fact, as a result of our approach we obtain that the three  particular cases above mentioned  are not exceptional values (with distinguishing properties) but just particular values in a more general situation.
Moreover  the above three Hamiltonians $\CaH_j$, $j=1,2,3$, obtained in \cite{SzumiMacPrz15} are the particular cases $(k_0=0, k_1=-1,k_2=0)$ of the following more general functions
$$
  H_{1}  =   {\smallonehalf}\,r^6\,\Bigl( p_r^2 +  \frac{p_{\phi}^2}{r^2} \Bigr) +  k_0 r^2 + r  \bigl( k_1 \cos\phi + k_2 \sin\phi \bigr)  \,, 
$$
$$
  H_{2}  =  {\smallonehalf}\,\frac{1}{r^2}\,\Bigl( p_r^2 +  \frac{p_{\phi}^2}{r^2} \Bigr) +  \frac{k_0}{r^2} + \frac{1}{r}\, \bigl( k_1 \cos\phi + k_2 \sin\phi \bigr) \,, 
$$
$$
  H_{3}  =  {\smallonehalf}\,r^4\,\Bigl( p_r^2 +  \frac{p_{\phi}^2}{r^2} \Bigr) +  \frac{k_0}{r^2} + \frac{1}{r}\bigl(k_1 \cos\phi + k_2 \sin\phi \bigr)  \,. 
$$

\item We obtain several families of superintegrable Hamiltonians with a  position dependent mass (pdm) of the form $m_n = r^{2 n}$  but with different  potentials $U(r,\phi)$. An important property is that these new potentials, that have also the form of a linear combination with coefficients $k_0$, $k_1$, and $k_2$, can be considered as the $m_n$-deformed versions of the Euclidean superintegrable potentials $V_a$ and $V_b$ (related with the harmonic oscillator), $V_c$ (related with the Kepler problem), and $V_d$ (also related with the Kepler problem), first obtained in \cite{FrMSUW65} and then studied by many authors (see \cite{Montreal} and references therein).

\end{itemize}

We close this Introduction with the following  comments.

First, the study of systems with a position dependent mass is a matter highly studied in these last years but, in most of cases, these studies are related with the problem of the quantization (because the problem of order in the  quantization of the kinetic term); the study presented in this paper is concerned with only the classical case and, although different, it has a close relation with the study presented in \cite{RanJmp15}.

Second,  quadratic superintegrability is a property very related with  Hamilton-Jacobi (H-J) multiple separability (Schr\"odinger separability in the quantum case) and this property is also true for  systems with a position dependent mass. 
This question (H-J separability approach to  systems with a pdm) was studied in \cite{RanJmp15} (in this case the pdm depends on a parameter $\kappa$) and more recently in \cite{Fordy16arXiv} (in this last case the pdm Hamiltonians studied were also related with those recently obtained  through a differential Galois group analysis in \cite{SzumiMacPrz15}).

Third,   the study of systems  admitting generalizations of the Laplace-Runge-Lenz vector \cite{LeachGor88}--\cite{Nikitin14} and the study of generalizations of the Kepler problem (Kepler-related problems with closed trajectories) are two (related) questions highly studied (see \cite{BizBor14} and references therein).
We will see in the next sections that some of the pdm Hamiltonians studied  in this paper are endowed with  integrals of motion rather similar to the Laplace-Runge-Lenz vector (this is also true for the above mentioned functions $H_{1}$ and $H_{2}$) and  therefore they belong to the family  of generalizated the Kepler problems.

\section{Superintegrability with quadratic constants of motion in the Euclidean plane } \label{Sec2}

We recall, in this Section, the existence  in the Euclidean plane of four two-dimensional potentials  $V_j$, $j=a,b,c,d$, that are superintegrable with quadratic integrals of motion.

\begin{enumerate}
\item[(a)]  The following potential, related with the harmonic oscillator, 
\begin{equation}   
  V_a =   {\smallonehalf}\,{\om_0}^2 (x^2 + y^2) + \frac{k_1}{x^2}  + \frac{k_2}{y^2}
\end{equation}
  is separable in Cartesian coordinates and polar coordinates. The constants of motion are the two one-dimensional energies and a third function related with the square of the angular momentum. 
  
\item[(b)]  The following potential, related with the harmonic oscillator, 
\begin{equation}   
  V_b =   {\smallonehalf}\,{\om_0}^2 (x^2 + 4 y^2) + \frac{k_1}{x^2}  + k_2 y 
\end{equation}
is separable in  Cartesian coordinates and  parabolic coordinates. 
The constants of motion are the two one-dimensional energies and a third function related with the Runge-Lenz vector. 

\item[(c)]  The following potential, related with the Kepler problem, 
\begin{equation}   
  V_c =  \frac{k_0}{\sqrt{x^2 + y^2}}  +  \frac{k_1}{y^2}  +  \frac{k_2\,x}{y^2 \sqrt{x^2 + y^2}}
\end{equation}   
is separable in  polar coordinates and  parabolic coordinates. 
The first constant of motion is the Hamiltonian itself and the other two,  $I_{c2}$ and $I_{c3}$, are related with the square of the angular momentum and the Runge-Lenz vector. 

\item[(d)]  The following potential, related with the Kepler problem, 
\begin{equation}   
V_d =  \frac{k_0}{\sqrt{x^2 + y^2}}   
  + k_1 \frac{ \bigl[\sqrt{x^2 + y^2} + x\bigr]^{1/2}}{ \sqrt{x^2 + y^2}}   
  + k_2 \frac{ \bigl[\sqrt{x^2 + y^2} - x\bigr]^{1/2} }{ \sqrt{x^2 + y^2}}
\end{equation}   
 is separable in  two different systems of parabolic coordinates and the  two constants of motion,  $I_{d2}$ and $I_{d3}$, are related with the Runge-Lenz vector.
\end{enumerate}

In the following sections we will study Hamiltonians with a pdm. 
We will prove that  the systems obtained in \cite{SzumiMacPrz15} are just particular cases of a much more general situation and we will present all the results making use of a notation that stress  the relation of the new Hamiltonians (to be denoted as $H_{nj}$. $j=a,b,c,d$) with the above mentioned  Euclidean systems.

\section{Position dependent mass, Killing vectors and Noether momenta}

A position  dependent mass $m_n = 1/r^{2 n}$ determines a kinetic Lagrangian $L_n=T_n$ and an  associated metric $ds_n^2$   given by  
\begin{equation}
 T_n=  {\smallonehalf}\,\frac{1}{r^{2 n}}\,\Bigl( v_r^2 +  r^2\,v_{\phi}^2 \Bigr)  \,,{\qquad}
 ds_n^2 =  \frac{1}{r^{2 n}}\,\Bigl( dr^2 +  r^2\,d{\phi}^2 \Bigr) \,. 
\end{equation}
This metric admits three symmetries; the invariance under rotations (generated by $X_J=\partial/\partial\phi$ ) and two other symmetries generated by the  Killing vectors $X_1$ and $X_2$ given by 
$$
 X_1 = r^n\,\Bigl(\cos(k_n\phi)\,\fracpd{}{r} + \frac{1}{r}\, \sin(k_n\phi) \,\fracpd{}{\phi} \Bigr)
  \,,{\qquad} 
 X_2 = r^n\,\Bigl(\sin(k_n\phi)\,\fracpd{}{r} - \frac{1}{r}\, \cos(k_n\phi) \,\fracpd{}{\phi} \Bigr)   \,,  
$$
where, for ease of the notation, we introduce $k_n$ for $k_n=n-1$. 
Every Killing vector $X$ determines an associated Noether momenta $P$ (so many Noether momenta as Killing vectors) that represents a constant of motion for the geodesic motion; 
so, in this case, we have the angular momentum $p_\phi=v_\phi/r^{2(n-1)}$ and the other two given by 
$$
  i(X_1)\,\te_L = \frac{1}{r^n}\,\Bigl( \cos(k_n\phi)\,v_r + r\sin(k_n\phi) \,v_\phi   \Bigr) 
\,,{\qquad} 
 i(X_2)\,\te_L = \frac{1}{r^n}\,\Bigl( \sin(k_n\phi)\,v_r - r \cos(k_n\phi) \,v_\phi   \Bigr) \,,
$$
where $\te_L$ is the Cartan 1-form 
$$
  \te_L = \bigl(\fracpd{L}{v_r}\Bigr)dr +  \bigl(\fracpd{L}{v_\phi}\Bigr)d\phi \,. 
$$
Making use of the Legendre transformation we obtain the kinetic Hamiltonian
$$
  H_n = T_n =  {\smallonehalf}\,r^{2 n}\,\Bigl( p_r^2 +  \frac{1}{r^2}\,p_{\phi}^2 \Bigr)  
$$
and the Hamiltonian expressions of the Noether momenta as linear functions of the canonical momenta 
$$
 P_{1} = r^n\,\bigl( p_r \cos(k_n\phi) + \frac{1}{r}\,p_\phi \sin(k_n\phi) \bigr) \,,  {\quad}
 P_{2} = r^n\,\bigl( p_r \sin(k_n\phi) - \frac{1}{r}\, p_\phi \cos(k_n\phi) \bigr) \,, 
$$
such that 
$$
  \{P_1\,,\, T_n\} = 0 \,,{\quad}   \{P_2\,,\, T_n\} = 0 \,,{\quad}   \{p_\phi\,,\, T_n\} = 0 \,.
$$

\section{Harmonic Oscillator  related Hamiltonians  }

In what follows we introduced potentials in the Lagrangian $L_n$ (Hamiltonian $H_n$)  in two steps. 
First central potentials ($V_{na}= 1/r^{2(n-1)}$ and $V_{nc}= r^{n-1}$) and then $\phi$-dependent new terms.
\subsection{Hamiltonian $H_{na}$}

 The first system to be studied with a position dependent mass $m_n=1/r^{2 n}$  is represented by a Hamiltonian with central potential $V_{na}= 1/r^{2 k_n}$
\begin{equation}
 H_{na}  =  T_n +  \frac{k_0}{r^{2 k_n}} \,,{\quad} k_n=n-1   \,,{\quad} n\ne 1 \,. 
\end{equation}
It is superintegrable with the following three constants of motion 
$$
 J_1  =   p_\phi     \,,{\qquad}  
 J_{11}  =   P_1^2  + 2\, \frac{k_0}{r^{2 k_n}} \bigl(\cos(k_n\phi)\bigr)^2   \,,{\qquad}  
 J_{22}  =   P_2^2  + 2 \,\frac{k_0}{r^{2 k_n}} \bigl(\sin(k_n\phi)\bigr)^2  \,, 
$$
that satisfy the following properties 
$$
 (i)\   dJ_1\,\wedge\,dJ_{11}\,\wedge\, dJ_{22}\ne 0 \,,{\quad} 
 (ii)\   \{J_{11}\,,\, J_{22}\} = 0 \,,{\quad} 
 (iii)\   H_{na} = \frac{1}{2}\bigl(J_{11} + J_{22} \bigr) \,. 
$$
A remarkable property is that the following function
$$
 J_{12}  =  J_{21} =   P_1 P_2  + 2\, \frac{k_0}{r^{2 k_n}} \cos(k_n\phi)  \sin(k_n\phi)
$$
is also a constant of motion. 
 These three integrals $\{J_{11},J_{22}, J_{12}\}$ can be considered as the three components $F_{ij}$, $i,j=1,2$, of a Fradkin tensor \cite{Frad65}. Because of this the Hamiltonian $H_{na}$ can be interpreted as representing an harmonic oscillator with a pdm $m_n=1/r^{2 n}$.
 
     In a similar way to what happens in the Euclidean case  \cite{FrMSUW65, Montreal}, the above Hamiltonian, that it has a central potential $V_{na}$, admits the addition of two non-central new terms preserving the quadratic superintegrability. In this case we have 
\begin{equation}
 H_{na}  =   T_n +  U_{na}(r,\phi) \,,  {\quad}    
 U_{na} =  \frac{k_0}{r^{2 k_n}} +  r^{2 k_n} \Bigl[\Bigl( \frac{k_1 }{\cos^2k_n\phi}  \Bigr) + \Bigl( \frac{k_2 }{\sin^2k_n\phi}  \Bigr)  \Bigr] \,, 
\end{equation}
where $k_0$, $k_1$, and $k_2$ are  arbitrary constants. 
The three independent constants of motion are 
$$
 J_{a1}  =  P_1^2  + 2\, \frac{k_0}{r^{2 k_n}} \bigl(\cos(k_n\phi)\bigr)^2 
 + 2 k_1 r^{2 k_n} \bigl(\sec(k_n\phi)\bigr)^2  \,,{\qquad}  
 J_{a2}  =  P_2^2  + 2 \,\frac{k_0}{r^{2 k_n}} \bigl(\sin(k_n\phi)\bigr)^2 
 + 2 k_2 r^{2 k_n} \bigl(\csc(k_n\phi)\bigr)^2  \,. 
$$
and
$$
 J_{a3} = p_\phi^2 +  2  \,\Bigl[\Bigl( \frac{k_1 }{\cos^2k_n\phi}  \Bigr) + \Bigl( \frac{k_2 }{\sin^2k_n\phi}  \Bigr)\Bigr]  \,. 
$$
 
  Starting with the central potential  $V_{na}= 1/r^{2 k_n}$ we can also construct the following Hamiltonian  \begin{equation}
 H_{na}'  =   T_n +  U_{na}'(r,\phi) \,,  {\quad}    
 U_{na}' =  \frac{k_0}{r^{2 k_n}} +  \frac{1}{r^{k_n}}\Bigl( k_1 \cos (k_n\phi)  +  k_2  \sin (k_n\phi)  \Bigr)  \,, 
\end{equation}
It has, in addition to the two quadratic constants $J_{a1}'$ and $J_{a2}'$, similar to $J_{a1}$ and $J_{a2}$, a linear in the momenta constant of motion 
$$
 J_{a3}' = 2 k_0 p_\phi + k_2 P_1 - k_1 P_2
$$
determined by an exact Noether symmetry of the Lagrangian $L_{na}' = T_n -  U_{na}'(r,\phi) $. 
Note that the Hamiltonian  $\CaH_3$  \cite{SzumiMacPrz15} mentioned in the introduction appears as the particular case $n=2$ of  $H_{na}'$.

\subsection{Hamiltonian $H_{nb}$}

Now we consider the pdm Hamiltonian  $ H_{nb}  =   T_n +  U_{nb}(r,\phi)$ 
where the potential $U_{nb}$ takes the form 
\begin{equation}
 U_{nb} =  \frac{k_0}{r^{2 k_n}} \Bigl( {\cos^2(k_n\phi}) + 4 \,{\sin^2(k_n\phi})  \Bigr)
 +  r^{2 k_n}  \Bigl( \frac{k_1 }{\cos^2 k_n\phi}  \Bigr) + \frac{k_2}{r^{k_n}} \sin(k_n\phi)   \,,{\quad} k_n=n-1  \,, 
\end{equation}
where $k_0$, $k_1$, and $k_2$ are  arbitrary constants. 
It is superintegrable with the following three independent  integrals of motion 
$$
 J_{b1}  =  P_1^2  + 2\, \frac{k_0}{r^{2 k_n}} \bigl(\cos(k_n\phi)\bigr)^2 
 + 2 k_1 r^{2 k_n} \bigl(\sec(k_n\phi)\bigr)^2  \,,{\qquad}  
 J_{b2}  =  P_2^2  + 8 \,\frac{k_0}{r^{2 k_n}} \bigl(\sin(k_n\phi)\bigr)^2 
 + \frac{2 k_2}{r^{k_n}} \sin(k_n\phi)   \,. 
$$
and 
$$
J_{b3}   =    P_1 p_\phi - \frac{k_0}{r^{3k_n}} \cos(k_n\phi) \sin(2 k_n\phi) + 
  k_1 r^{k_n} \bigl(\sec^3 (k_n\phi )\,\sin(2 k_n\phi) \bigr) 
 -\frac{k_2}{2  r^{2 k_n}} \cos^2(k_n\phi)  \,.  
$$

\section{Kepler related Hamiltonians  }

\subsection{Hamiltonian $H_{nc}$}

Now we consider a Hamiltonian with a position dependent mass $m_n=1/r^{2 n}$ and a central potential $V_{nc} = r^{n-1}$
\begin{equation}
 H_{nc}  =  T_n +  k_0 r^{n-1}  \,,{\quad} n\ne 1 \,. 
\end{equation}
It is superintegrable with the following three constants of motion 
$$
 J_1 = p_\phi   \,,{\qquad}  
  J_{2} =   P_2 p_\phi - k_0 \cos(k_n\phi)   \,, {\qquad}  
  J_{3} =   P_1 p_\phi + k_0 \sin(k_n\phi)   \,,   
$$
It is clear that $J_{2}$ and $J_{3}$ are quite similar to the two components of a two-dimensional Runge-Lenz vector.   Because of this the Hamiltonian $H_{nc}$ can be interpreted as representing a Kepler system with a pdm $m_n=1/r^{2 n}$.

There are three different ways of modifying the potential $V_{nc}$ by introducing  additional $\phi$-dependent terms in such a way that the superintegrability is preserved. 
In the two first cases only one of the two Runge-Lenz-like constants is preserved; 
in the third case both two Runge-Lenz-like constants are preserved (but then the integral $J_1$ disappears). 

\begin{itemize}

\item [(c1)]   The following Hamiltonian 
\begin{equation}
 H_{nc1}  =  T_n +  U_{nc1}(r,\phi) \,,
{\quad}    
U_{nc1} =  k_0 r^{n-1} + r^{2 k_n} \Bigl[\Bigl( \frac{k_1 }{\sin^2k_n\phi}  \Bigr) + k_2\Bigl( \frac{\cos k_n\phi }{\sin^2k_n\phi}  \Bigr)  \Bigr] \,, 
\end{equation}
 has  (in addition to the Hamiltonian itself) two functionally independent first integrals of the second order in the momenta
$$
    dJ_{c2}\,\wedge\,dJ_{c3}\,\wedge\,dH_{nc1} \ne 0 \,,{\quad} 
   \{J_{c2}\,,\, H_{nc1} \} = 0 \,,{\quad} 
    \{J_{c3}\,,\, H_{nc1} \} = 0 \,, 
$$
given by
\begin{eqnarray*}   
 J_{c2}  &=&   p_\phi^2 +  2  \,\Bigl[\Bigl( \frac{k_1 }{\sin^2k_n\phi}  \Bigr) 
+ k_2\Bigl( \frac{\cos k_n\phi }{\sin^2k_n\phi}  \Bigr)  \Bigr]    \,,  \cr
 J_{c3}  &=&   P_2 p_\phi - k_0 \cos(k_n\phi) -  2  k_1 r^{k_n} \bigl(\csc k_n\phi\cot k_n\phi \bigr) 
 -  k_2 r^{k_n} \bigl(\csc^2k_n\phi + \cot^2k_n\phi \bigr)   \,.  
 \end{eqnarray*}   
 
\item [(c2)]   The Hamiltonian  
\begin{equation}
 H_{nc2}  =   T_n +  U_{nc2}(r,\phi) \,,
{\quad}    
U_{nc2} =k_0 r^{n-1} + r^{2 k_n} \Bigl[\Bigl( \frac{k_1 }{\cos^2k_n\phi}  \Bigr) 
+ k_2\Bigl( \frac{\sin k_n\phi }{\cos^2k_n\phi}  \Bigr)  \Bigr] \,, 
\end{equation}
is similar to the previous one $H_{nc1}$ but in this case is  the existence of the second Runge-Lenz integral what is preserved 
\begin{eqnarray*}   
 J_{c2}  &=&   p_\phi^2 +  2  \,\Bigl[\Bigl( \frac{k_1 }{\cos^2k_n\phi}  \Bigr) 
+ k_2\Bigl( \frac{\sin k_n\phi }{\cos^2k_n\phi}  \Bigr)  \Bigr]    \,,  \cr
 J_{c3}  &=&   P_1 p_\phi + k_0 \sin(k_n\phi) +  2  k_1 r^{k_n} \bigl(\sec k_n\phi\tan k_n\phi \bigr) 
 +  k_2 r^{k_n} \bigl(\sec^2k_n\phi + \tan^2k_n\phi \bigr)   \,.  
 \end{eqnarray*}   
\end{itemize}

\subsection{Hamiltonian $H_{nd}$}  

   The Hamiltonian  
\begin{equation}
  H_{nd}  =   T_n +  U_{nd}(r,\phi)  
\,,{\quad}  
 U_{nd} = k_0  r^{n-1} + r^{k_n/2}  \bigl( k_1 \cos k_n(\phi/2) + k_2 \sin k_n(\phi/2) \bigr)  \,, 
\end{equation}
that generalizes the Hamiltonians $\CaH_1$ and $\CaH_2$ obtained in \cite{SzumiMacPrz15} and mentioned in the Introduction (they correspond to $n=3$ and $n=-1$).   
It possesses the following two independent constants of motion $J_{d2}$ and $J_{d3}$ 
  \begin{eqnarray*}   
 J_{d2}  &=&  P_1 p_\phi - k_0 \cos(k_n\phi) +   \frac{k_1}{r^{k_n/2}} \bigl(\sin k_n\phi\sin k_n(\phi/2)\bigr)   -   \frac{k_2}{r^{k_n/2}} \bigl(\sin k_n\phi\cos k_n(\phi/2) \bigr)   \,,  \cr
 J_{d3}  &=&  P_2 p_\phi + k_0 \sin(k_n\phi)   +   \frac{k_1}{r^{k_n/2}} \bigl(\cos k_n\phi\sin k_n(\phi/2) \bigr)   -   \frac{k_2}{r^{k_n/2}} \bigl(\cos k_n\phi\cos k_n(\phi/2) \bigr)  \,.
  \end{eqnarray*}   
Both are of Runge-Lenz type.

\section{Complex functions and Superintegrability  } \label{Sec6}

We mention in the Introduction the existence of different approaches (Noether symmetries, Hidden symmetries, Lax pairs formalism, bi-Hamiltonian structures, H-J separability) for the study of Liouville integrability (or superintegrability). 
Now in this section we study the superintegrability of two of the Hamiltonians 
($H_{na}' $ related to the harmonic oscillator and $H_{nd}$ related to the Kepler problem) already studied in the previous section but now making use of a rather different approach. 
The main idea is that the superintegrability can be related with the existence of certain complex functions with interesting Poisson brackets properties. 
This complex functions formalism has been recently studied in \cite{CaRa16Sigma} for the Kepler problem in the Euclidean plane.

\subsection{Hamiltonian $H_{na}' $}

Let us first introduce   the following real functions 
\begin{eqnarray*}   
 M_{n1} &=&    r^{2 k_n}\,\bigl(r^2 p_r^2 -  p_\phi^2 \bigr) 
 + \frac{2k_0}{r^{2 k}} + \frac{2}{r^{k_n}} \bigl(k_1 \cos(k_n\phi) + k_2 \sin(k_n\phi) \bigr) \,, \cr
 M_{n2} &=&    2 r^{2 n-1} p_r p_\phi  + \frac{2}{r^{k_n}} \bigl(k_1 \sin(k_n\phi) - k_2 \cos(k_n\phi)\bigr) \,, 
\end{eqnarray*}
and  
$$
  N_{\phi1} =  \cos (2k_n\phi)   \,,{\quad}   N_{\phi2} =  \sin (2k_n\phi) \,.
$$
Then  if we denote by $M_n$ and $N_{\phi}$ the   complex functions 
$$
  M_n = M_{n1} + i\,M_{n2}   \,,{\qquad}  N_{\phi} = N_{\phi1} + i\,N_{\phi2}\,, 
$$
we have 
$$
  \frac{d}{d t}\,M_{n}  = \{M_{n}\,,\,H_{na}'\} = i\, 2\,{\lambda_n} M_{n}
\,,{\qquad} 
  \frac{d}{d t}\,N_{\phi}  = \{N_{\phi}\,,\,H_{na}'\} = i\, 2\,{\lambda_n} N_{\phi} \,,  
$$
where the common factor $\lambda_n$ is   given by  
$$
  \lambda_n =  (n-1) r^{2 k_n}\, p_\phi \,.
$$
This means that the function constructed by coupling $M_n$ with $N_{\phi}$ is a constant of motion. 
This result is presented in the following proposition.

\begin{proposition}  \label{proposition1} 
Let us consider  the following Hamiltonian with a position dependent mass $m=r^{2n}$
$$
 H_{na}'  =   T_n +  U_{na}'(r,\phi) \,,  {\quad}    
 U_{na}' =  \frac{k_0}{r^{2 k_n}} +  \frac{1}{r^{k_n}}\Bigl( k_1 \cos (k_n\phi)  +  k_2  \sin (k_n\phi)  \Bigr)  \,, 
$$
Then, the complex function $J_{23}$ defined as 
$$
  J_{23} = M_n  N_{\phi}^{*} 
$$
is a quadratic (complex) constant of  motion. 
\end{proposition} 
Of course $J_{23}$ determines two real first-integrals 
$$
 J_{23} = J_2 + i\, J_3 \,,{\quad}
 \bigl\{J_2\,, H_{na}' \bigr\} =  0 \,,{\quad}
 \bigl\{J_3\,, H_{na}' \bigr\} =   0 \,. 
$$
whose coordinate expressions turn out to be 
 \begin{eqnarray*}   
   J_2 &=&    r^{2(n-1)}  \Bigl( ( r^2  p_r^2 -  p_\phi^2) \cos(2k_n\phi)   + (2 r  p_r p_\phi  )\sin(2k_n\phi) \Bigr)   + \frac{2}{r^{2k_n}} k_0 \cos(2k_n\phi)  
   +  \frac{2}{r^{k_n}} \bigl(k_1 \cos(k_n\phi) -   k_2 \sin(k_n\phi) \bigr)  \,,  \cr
 J_3  &=&    r^{2(n-1)}  \Bigl( ( r^2  p_r^2 -  p_\phi^2) \sin(2k_n\phi)   - (2 r  p_r p_\phi  )\cos(2k_n\phi) \Bigr)   + \frac{2}{r^{2k_n}} k_0 \sin(2k_n\phi)  
 +  \frac{2}{r^{k_n}}\bigl(k_1 \sin(k_n\phi) +   k_2 \cos(k_n\phi)\bigr)    \,.
\end{eqnarray*}
 Concerning the linear constant of motion $J_{a3}'$ (obtained from an exact Noether symmetry), it determines the following Poisson brackets wit $J_2$ and $J_3$ 
 $$
 \{J_{a3}'\,,\,J_2\} = 4 (n-1)\bigl(k_0 J_3 +  k_1 k_2 \bigr)
  \,, {\quad}
 \{J_{a3}'\,,\,J_3\} = -\, 2\,(n-1)\bigl(2 k_0 J_2 +  k_1^2 - k_2^2\bigr)  \,. 
 $$

\subsection{Hamiltonian $H_{nd}$ }

Let us denote by  $A_{nj}$ and $N_{\phi j}$, $j=1,2$,  the following real functions 
$$
 A_{n1} =  r^{n-1}\,p_\phi^2 + k_0  \,,{\quad}  
 A_{n2} =  \frac{1}{r^{k_n/2}} \Bigl( r^{m_n}\,p_r p_\phi + k_1 \sin (k_n/2)\phi - k_2 \cos  (k_n/2)\phi  \Bigr)  
 \,,\  m_n= {\smallonehalf}(3 n - 1)\,,
$$
and 
$$
  N_{\phi1} =  \cos k_n\phi   \,,{\quad}   N_{\phi2} =  \sin k_n\phi \,.
$$
Then we have the following properties
$$
\textrm{(i)}{\qquad}  
\frac{d}{d t}\,A_{n1}  = \{A_{n1}\,,\,H_{nd}\} =  (n-1)\,\lambda_n\,A_{n2}
\,,{\qquad} 
  \frac{d}{d t}\,A_{n2}  = \{A_{n2}\,,\,H_{nd}\} = -\, (n-1)\,\lambda_n\,A_{n1} \,,
$$
$$
\textrm{(ii)}{\qquad}   \frac{d}{d t}\,N_{\phi1}  = \{N_{\phi1}\,,\,H_{nd}\} = -\, (n-1)\,\lambda_n\,N_{\phi2}
\,,{\qquad} 
  \frac{d}{d t}\,N_{\phi2}  = \{N_{\phi2}\,,\,H_{nd}\} = \, (n-1)\,\lambda_n\,N_{\phi1} \,,
$$
where $\lambda_n$ denotes the following function 
\begin{equation}
  \lambda_n = r^{2(n-1)}\,  p_\phi   \,. 
\end{equation}
Therefore, the two complex functions $A_n$ and $N_\phi$ defined as 
$$
 A_n = A_{n1} + i\, A_{n2} \,,{\quad} N_\phi = N_{\phi1} + i\,N_{\phi2} \,,
$$
satisfy the following Poisson bracket properties 
$$
   \{A_n\,,\,H_{nd}\} = -\,i\,(n-1)\, \lambda_n\,A_{n}  \,,{\qquad} 
   \{N_{\phi}\,,\,H_{nd}\} = i\,(n-1)\, \lambda_n\,N_{\phi} \,,
$$
and  consequently the Poisson bracket of  the complex function $A_n N_\phi$ with the Kepler-related  Hamiltonian $H_{nd}$ vanishes 
\begin{eqnarray*}   
  \{A_n \,N_\phi \,,\,H_{nd}\}  &=&   \{A_n  \,,\,H_{nd}\}\,N_\phi
   +   A_n\,\{N_\phi\,,\,H_{nd}\}    \cr
    &=&   (n-1)\bigl(-\,i\, \lambda_n\,A_n\bigr)\,N_\phi 
   +   (n-1) A_n \bigl( i\, \lambda_n\,N_{\phi}\bigr)  = 0  \,. 
\end{eqnarray*}

We  can summarize this result in the following proposition. 

\begin{proposition}  \label{proposition1}
Let us consider  the Kepler-related   Hamiltonian  $H_{nd}$ with pdm $m_n = 1/r^{2 n}$
$$
  H_{nd}  =   T_n +  U_{nd}(r,\phi)  
\,,{\quad}  
 U_{nd} = k_0 r^{k_n} + r^{k_n/2}  \bigl( k_1 \cos(k_n/2)\phi + k_2 \sin(k_n/2)\phi \bigr)  \,, 
$$
Then, the complex function $J_{23}$ defined as 
$$
  J_{23} = A_n N_\phi
$$
is a quadratic (complex) constant of  motion. 
\end{proposition}

Of course $J_{23}$ determines two real first-integrals 
$$
 J_{23} = {\Re}(J_{23}) + i\, {\Im}(J_{23}) \,,{\quad}
 \bigl\{{\Re}(J_{23})\,, H_{nd}\bigr\} =  0 \,,{\quad}   
 \bigl\{{\Im }(J_{23})\,, H_{nd}\bigr\} =   0 \,, 
$$
whose coordinate expressions turn out to be 
$$
 {\rm Re}(J_{23})  =  J_{d2} \,,{\qquad}
 {\rm Im}(J_{23})  =  J_{d3}   \,.
$$
That is, the two real functions ${\Re}(J_{23})$ and ${\Im}(J_{23})$  are just the two components of the pdm-version of the two-dimensional Laplace-Runge--Lenz vector.  

Summarizing, we have got two interesting properties. 
First, the superintegrability of the pdm-deformed version $H_{nd} $ of the Kepler problem is directly related with the existence of two complex functions ($ A_n$ and $N_\phi$) whose Poisson brackets with the Hamiltonian $H_{nd}$ are proportional, with a common complex factor, to themselves; 
and second,  the two components of the pdm-deformed version of the Laplace-Runge-Lenz vector appear as the real and imaginary parts of the complex first-integral of motion. 
Remark that $ N_\phi$ is a complex function of constant modulus one, while the modulus of $ A_n$ is a polynomial of degree four in the momenta that is just the sum of the squares of $J_{d2}$ and $J_{d3}$ 
$$
 A_n\,A_n^* = J_{d2}^2 +  J_{d3}^2   \,.
$$

\section{Final comments }

We have studied the superintegrability of Hamiltonian systems with a pdm $m_n=r^{2 n}$, $n\ne 1$,  and we have proved that the particular Hamiltonians previously obtained in [1] are just very particular cases of the systems here obtained. We have made use of the properties of Killing vectors as the starting point of our approach, and we have proved that the Hamiltonians so obtained can be considered as pdm-deformations of the classical Euclidean superintegrable systems with potentials $V_a$ and $V_b$ (related with the harmonic oscillator), and $V_c$ and $V_d$ (both related with the Kepler problem). 
This result clearly reinforce the importance of these four potentials since, although defined in an Euclidean geometry,  
they are directly related with superintegrable systems with a nonEuclidean metric (this close relation between superintegrable Hamiltonians with and without pdm   was already considered in \cite{RanJmp15}). 

Integrability and superintegrability  on spaces of constant and nonconstant curvature is a matter recently studied by several authors (see, e.g. \cite{KalKrMiPog02, KalKrWint02, BurgosPhysD08, BurgosSigma11} and references therein). 
Nevertheless, in differential geometric terms a pdm global factor means that the configuration space $Q$ is endowed with a conformal metric (a nonEuclidean space but with conformal equivalence to the Euclidean one) and in this case we have the additional property that the pdm is a function dependent only of the radial variable.
 This is probably the main reason for the existence of a so close relation between the pdm Hamiltonians we have obtained and the four Euclidean superintegrable systems  mentioned in Section \ref{Sec2}. 
Moreover, and concerning that existence of curvature, we recall that in two dimensions the Riemann tensor  $R_{abcd}$ only has one independent component which can be taken $R_{1212}$ 
$$
 R_{1212} =  \frac{1}{2}\,\bigl(\partial_2\partial_1g_{21} - \partial_2^2g_{11} + \partial_1\partial_2g_{12} - \partial_1^2g_{22} \bigr) 
 - g_{ef}\bigl( \Ga_{11}^e\Ga_{22}^f   -   \Ga_{12}^e\Ga_{21}^f  \bigr) \,. 
$$
In this case (with $g_{11} = 1/r^{2n}$ and  $g_{22} = 1/r^{2n-2}$) the result is $ R_{1212} = 0 $. So the configuration space for the Hamiltonian $H_n$ (that is, $Q=\IR^2$ with the line element $ds_n^2$)  is in fact a flat manifold.

We finalize with the following  questions for future work. 

\begin{itemize}
\item{} It is natural to suppose the existence of superintegrable systems with a position dependent mass but with higher order constants of motion.  We recall that the Euclidean potentials $V_a$  and $V_c$ admit two generalizations  
$$
  V_{ttw}(r,\phi)  =  {\smallonehalf}\,{\om_0}^2 r^2 +  \frac{1}{2\,r^2}\,\Bigl(   \frac{\alpha}{\cos^2(m\,\phi)}  +  \frac{\beta}{\sin^2(m\,\phi)} \Bigr) \,,   
$$
$$
  V_{pw}(r,\phi)  =  -\, \frac{g}{r}  +  \frac{1}{2\,r^2}\,\Bigl(   \frac{\alpha}{\cos^2(m\phi)} + \frac{\beta}{\sin^2(m\phi)} \Bigr) \,,   \label{VpwE2}
$$
that are superintegrable but with higher order constants of motion \cite{TTW09, TTW10, PostWint10}.  
So,  the existence of superintegrable systems similar to these two Euclidean systems but with a pdm of the form $m_j$ is a matter to be studied. 
The higher surerintegrability of the potentials $V_{ttw}$ and $V_{pw}$ has been studied making use of different techniques (at both the classical and the quantum levels);  here we point out the existence of a method that make use of products of complex functions \cite{Ra12JPaTTW,Ra13JPaPW}; probably this method can also be applied to the study of the $m_j$-dependent case.

\item{} Concerning the complex functions formalism presented in Section  \ref{Sec6}, we mention that it was proved in \cite{CaRa16Sigma} that it is related with the existence of quasi-bi-Hamiltonian  structures. 
So the existence of these structures (bi-Hamiltonian or  quasi-bi-Hamiltonian) for systems with a pdm is also a matter to be studied making use of the properties of these complex functions. 

\item{}  Finally, the study of quantum systems with a pdm is a matter highly studied in these last years. First the quantization of these systems is not an easy matter (because the problem of order in the  quantization of the kinetic term) and second,  it seems that some of these pdm systems belong to the family of Hamiltonians with an exactly solvable Schr\"odinger equation. 
Therefore, the quantum study of all these pdm Hamiltonians is also an interesting  matter to be studied.

 \end{itemize}

\section*{Acknowledgments}

This work has been  supported by the research projects MTM--2012--33575 (MICINN, Madrid)  
and DGA-E24/1 (DGA, Zaragoza). 

{\small
       }

\end{document}